\documentclass[aps,floats,prd,nofootinbib,twocolumn]{revtex4}

\usepackage{amssymb}
\usepackage{amsmath}

\usepackage{graphicx}
\usepackage{graphics}
\usepackage{dcolumn}
\usepackage{color}
\usepackage{fancyhdr} 
\usepackage{graphicx}

\usepackage{subfig}
\usepackage{bibshortcuts}

\usepackage[utf8]{inputenc}

\usepackage[justification=centerlast, format=plain, labelfont=bf]{caption}

\def\VEV#1{\left\langle #1 \right\rangle}

\newcommand{\be}{\begin{equation}}
\newcommand{\ee}{\end{equation}}
\newcommand{\ba}{\begin{align}}
\newcommand{\ea}{\end{align}}

\begin{document}
	
	\title{Constraints on Dark Matter-Baryon Scattering from the Temperature Evolution of the Intergalactic Medium}
	
	\author{Julian B.~Mu\~noz\footnote{Electronic address: \tt julianmunoz@fas.harvard.edu}
	} 
	\affiliation{Department of Physics and Astronomy, Johns
		Hopkins University, Baltimore, MD 21218,}
	\affiliation{Department of Physics, Harvard University, Cambridge, MA 02138}
	\author{Abraham Loeb}
	\affiliation{Astronomy Department, Harvard University, 60 Garden St., Cambridge, MA 02138}
	
	\date{\today}

\begin{abstract}
	
The thermal evolution of the intergalactic medium (IGM) can serve as a sensitive probe of cosmological heat sources and sinks.
We employ it to limit 
interactions between dark matter and baryons.
After reionization the IGM temperature is  set by the balance between photoheating and adiabatic cooling. 
We use measurements of the IGM temperature from Lyman-$\alpha$-forest data to constrain the cross-section $\sigma$ between dark matter and  baryons, finding $\sigma < 10^{-20}$ cm$^2$ for dark-matter masses $m_\chi\leq 1$ GeV. 
This provides the first direct constraint on scattering between dark matter and baryons at redshift $z\sim5$.
\end{abstract}

\maketitle

\section{Introduction}

%Composing the majority of the matter in our Universe, cold dark matter presents no visible signs of interactions with baryons.

The majority of the matter in our Universe is dark, clustering under the force of gravity but otherwise not interacting with regular baryonic matter~\cite{Ade:2015xua,Alam:2016hwk,Silk:2016srn}.
The nature of this dark matter (DM) remains a mystery, with candidates for its composition ranging from scalar fields~\cite{Preskill:1982cy,Peebles:2000yy} to weakly interacting massive particles (WIMP)~\cite{Jungman:1995df,Bertone:2004pz}, or primordial black holes~\cite{Carr:1974nx,Bird:2016dcv,Brandt:2016aco,Ali-Haimoud:2016mbv}.

Most cosmological observables within our reach can be explained by entirely sterile dark matter---coupling only gravitationally to baryons.
However, a small amount of dark-matter--baryon interaction is possible,
and may in fact be favored to resolve some structure-formation puzzles~\cite{Moore:1997jy,Kamionkowski:1999vp,2010arXiv1009.4505B,Weinberg:2013aya}.
Moreover, interactions may be key to understanding why the dark-matter and baryonic energy densities are comparable in magnitude~\cite{Kaplan:1991ah,Allahverdi:2013mza}. 
Different dark-matter candidates can present distinct interactions with baryons.
It is then clear that measuring (or constraining) interactions between baryons and dark matter can be a powerful probe into the nature of dark matter.
In this paper we
propose a novel probe for these interactions,
involving the use of the intergalactic medium (IGM) as a thermostat.

A new interaction vertex between dark-matter and baryons can give rise to three avenues for detection.
First, baryons can annihilate and produce dark-matter particles, which could be detected in particle colliders~\cite{Goodman:2010ku}, such as LEP~\cite{Fox:2011fx} and the LHC~\cite{Aaboud:2016tnv}.
Second, in the converse process dark matter can annihilate and heat up the baryons~\cite{Galli:2009zc,Cirelli:2009bb}, or produce gamma-rays~\cite{TheFermi-LAT:2017vmf}.
Last, interactions cause dark matter and baryons to scatter off each other.
Direct-detection experiments have been able to severely constrain the scattering between dark matter and baryons for dark-matter particles heavier than $\sim$ 10 GeV~\cite{Amole:2016pye,Tan:2016zwf,Akerib:2017kat}.
Additionally, scattering allows baryons to transfer pressure to the dark matter, hampering structure formation and leaving an imprint on the cosmic microwave background (CMB) and on the Lyman-$\alpha$ forest~\cite{Dvorkin:2013cea}.

Here, we will focus on the effect of scattering on the thermal history of the IGM.
In Refs.~\cite{Tashiro:2014tsa,Munoz:2015bca} this effect was studied during the dark ages ($z\sim 30$), when the gas temperature could be measured indirectly through the 21-cm line. 
This effect also produces spectral distortions in the CMB, as explored in Ref.~\cite{Ali-Haimoud:2015pwa}.
Here we propose using gas-temperature measurements of the IGM to probe the scattering cross-section of dark-matter with baryons.
We focus on the era after hydrogen reionization but before the second reionization of helium, spanning the redshifts $z \sim 4-15$.
In this era the temperature of the IGM gas is set by the equilibrium between photoheating and adiabatic plus Compton cooling~\cite{Lidz:2014jxa}.
We will show how interacting with dark matter can significantly lower the temperature of the baryons. 
In particular, we use the gas-temperature measurements of Refs.~\cite{Becker2011,Bolton2012} to show that dark-matter--baryon cross sections $\sigma$ larger than $\sigma \approx 10^{-20}$ cm$^2$, for dark-matter masses below $1$ GeV, are disfavored by the data.
This is complementary to the constraints from Refs.~\cite{Dvorkin:2013cea} and \cite{Ali-Haimoud:2015pwa}, valid only for $m_\chi\gg 1$ GeV and $m_\chi\lesssim 100$ keV, respectively.

The outline of this paper is as follows. In Section~\ref{sec:Thermalization} we describe the heating due to dark-matter--baryon interactions. Later, in Section~\ref{sec:IGMTemperature} we find the evolution of the IGM temperature including interactions; and with it we constrain the scattering cross-section in Section~\ref{sec:Results}. 
We conclude in Section~\ref{sec:Conclusions}.

\section{Thermalization Rate}
\label{sec:Thermalization}

We begin with a non-relativistic system composed of dark-matter particles of mass $m_\chi$ and velocity $\mathbf v_\chi$,
and baryons of mass $m_b$ and velocity $\mathbf v_b$. In the center of mass (CM) of this system an elastic collision
does not alter the magnitude of the velocity of either particle, only their direction. 
We can thus parametrize the final velocity of the dark-matter particle as
\be
\mathbf v_\chi^{f,\rm CM} = v_\chi^{i,\rm CM} \hat n,
\ee
where $\hat n$ is a unit vector, and $v_\chi^{i,\rm CM}$ is the magnitude of the initial dark-matter velocity in the CM frame. The momentum transfer per collision is thus~\cite{Dvorkin:2013cea},
\be
\Delta \mathbf p_\chi = - \Delta \mathbf p_b = \dfrac{m_b\,m_\chi}{m_\chi + m_b} \left [ \left| \mathbf v_\chi - \mathbf v_b \right| \hat n - (\mathbf v_\chi - \mathbf v_b ) \right],
\ee
which is a (non-relativistic) Galilean invariant, and thus does not change when considered in the rest-frame of the baryons.
Therefore, the energy transferred to each baryon per interaction is given by
\be
\Delta E_b = \mathbf v_{\rm CM} \cdot \Delta \mathbf p_b = - \Delta E_\chi,
\label{eq:DEb}
\ee
where the center-of-mass velocity is $\mathbf v_{\rm CM} \equiv (m_b \mathbf v_b +m_\chi \mathbf v_\chi )/(m_b+m_\chi)$.

We now assume that both the baryonic and dark-matter particles behave as a fluid,
which can be described by a Maxwell-Boltzmann velocity distribution, 
\be
f_X (v)= (2\pi u_X^2)^{3/2} e^{-v^2/(2u_X^2)},
\label{eq:Boltzmann}
\ee
where $u_X^2\equiv T_X/m_X$, $X$ denotes the fluid ($b$ for baryons and $\chi$ for dark matter) with temperature $T_X$ and particle mass $m_X$, and we use units in which both the Boltzmann constant $k_B$, and the speed of light, $c$, are unity.
For simplicity we will only consider scattering off protons, so that $m_b = m_p = 0.938$ GeV, and we will refer to protons as baryons and gas indistinguishably.
The rate of collisions with recoil in the direction $\hat n$ for any given baryon is $|\mathbf v_\chi - \mathbf v_b| n_\chi d\sigma/d\hat n$, where $d \sigma$ is the differential cross section, and $n_\chi$ is the number density of the dark-matter particles, which act as targets.
During the redshift range of interest, the bulk relative velocity $\mathbf V_{\chi b} = \VEV{\mathbf v_\chi - \mathbf v_b}$ between baryons and dark matter is negligible, so we will ignore it. Thus, the heating rate of the baryon fluid is
\be
\dot Q_b = n_\chi \int d^3 \mathbf v_\chi  f_\chi \int d^3 \mathbf v_b f_b |\mathbf v_\chi - \mathbf v_b| \int d\hat n \dfrac{ d \sigma}{ d \hat n}  \Delta E_b (\hat n),
\ee
where the innermost integral can be recast as
\be
\int d\hat n \dfrac{ d \sigma}{ d \hat n} \Delta E_b (\hat n) = - \bar \sigma(|\mathbf v_\chi - \mathbf v_b|) \dfrac{m_b m_\chi}{m_b + m_\chi} \mathbf v_{\rm CM} \cdot (\mathbf v_\chi - \mathbf v_b),
\ee
by using Eq.~\eqref{eq:DEb} and
\be
\bar \sigma = \int d\!\cos\theta (1-\cos\theta) \dfrac{d\sigma}{d\cos\theta} \ .
\ee

We can rewrite the velocity integrals in terms of two new variables,
\ba
\mathbf v_p &\equiv \dfrac{u_b^2 \mathbf v_\chi +u_\chi^2 \mathbf v_b}{u_\chi^2 + u_b^2}, \qquad \rm and \nonumber \\
\mathbf v_m &\equiv \mathbf v_\chi -\mathbf v_b,
\end{align}
finding that
\be
\int d^3 \mathbf v_\chi  f_\chi \int d^3\mathbf  v_b f_b = \int d^3 \mathbf v_p  f_p \int d^3 \mathbf v_m f_m,
\ee
where $f_p$ and $f_m$ are Boltzmann distribution functions, as in Eq.~\eqref{eq:Boltzmann}, with widths $u_p^{-2} = u_b^{-2} + u_\chi^{-2}$ and $u_m^2 = u_b^2 + u_\chi^2$, respectively. 
We decompose the CM velocity as~\cite{Munoz:2015bca}
\be
\mathbf v_{\rm CM} = \mathbf v_p + \dfrac{(T_\chi-T_b)}{u_m^2\,(m_\chi+m_b)} \mathbf v_m,
\ee
and noticing that the integral over the variable $v_p$ is trivial, as interactions depend exclusively on $v_m$, we find
\be
\dot Q_{b,\chi} = -\dfrac{\rho_\chi \mu_b (T_\chi-T_b)}{u_m^2 \, (m_\chi+m_b)^2}\int d^3\mathbf  v_m  f_m  \, v_m^3 \bar \sigma(v_m),
\label{eq:Qdotint}
\ee
where $\mu_b$ is the mean molecular weight of the baryons, and
$\rho_i$ is the average energy density of species $i$ (either baryons or dark matter), given by $\rho_i(z) = \Omega_i \rho_{\rm crit} (1+z)^3$, where we set $\Omega_b = 0.045$ and $\Omega_\chi = 0.24$, and $ \rho_{\rm crit}=3H_0^2/(8\pi G)$ is the critical density. 

We parametrize 
the scattering cross section between dark matter and baryons as a power-law of their relative velocity~\cite{Dvorkin:2013cea},
\be
\bar \sigma(v_m) = \sigma_n v_m^n,
\label{eq:sigman}
\ee
where the index $n$ describes different interaction models, such as millicharged dark matter ($n=-4$) \cite{Cline:2012is}, and heavy-mediator interactions ($n=0$) \cite{Chen:2002yh}.
Under this assumption, we can evaluate Eq.~\eqref{eq:Qdotint} to find
the baryonic heating rate
\be
\dot Q_{b,\chi} = \Gamma_{b,\chi} (T_\chi - T_b),
\label{eq:Qdotb}
\ee
where the thermalization interaction rate is defined to be
\be
\Gamma_{b,\chi} \equiv \dfrac{\rho_\chi \mu_b \sigma_n}{(m_\chi+m_b)^2} \left(\dfrac{T_b}{m_b}+\dfrac{T_\chi}{m_\chi}\right)^{\frac{n+1} 2} \dfrac{2^{\frac {5+n} 2} \Gamma\left( 3+ \dfrac{n}{2} \right)}{\sqrt \pi}.
\label{eq:Gammabchi}
\ee

Notice that in the presence of a relative velocity between baryons and dark matter, these fluids will tend to reach both thermal and mechanical equilibrium, which can generate positive heat for both species~\cite{Munoz:2015bca}. The mechanical effect is negligible during the epoch we study, as the relative velocity---of order 0.1 km/s---is a factor $\sim 100$ smaller than the thermal velocity of baryons with a temperature of $\sim10^4$ K, so we safely neglect it.

\section{IGM Temperature}
\label{sec:IGMTemperature}

We next aim to find how the coupling in Eq.~\eqref{eq:Gammabchi} affects the evolution of the IGM temperature.
The baryonic gas, composed of hydrogen and helium, was first ionized as a consequence of star formation, with the bulk of the ionization occurring by $z\sim 7$~\cite{Adam:2016hgk}.
During this time the gas temperature is roughly determined by the equilibrium between ionization photoheating and cooling from both adiabatic expansion and Compton scattering with CMB photons.

\begin{figure*}[hbtp!]
	\centering
	\includegraphics[width=0.7\textwidth]{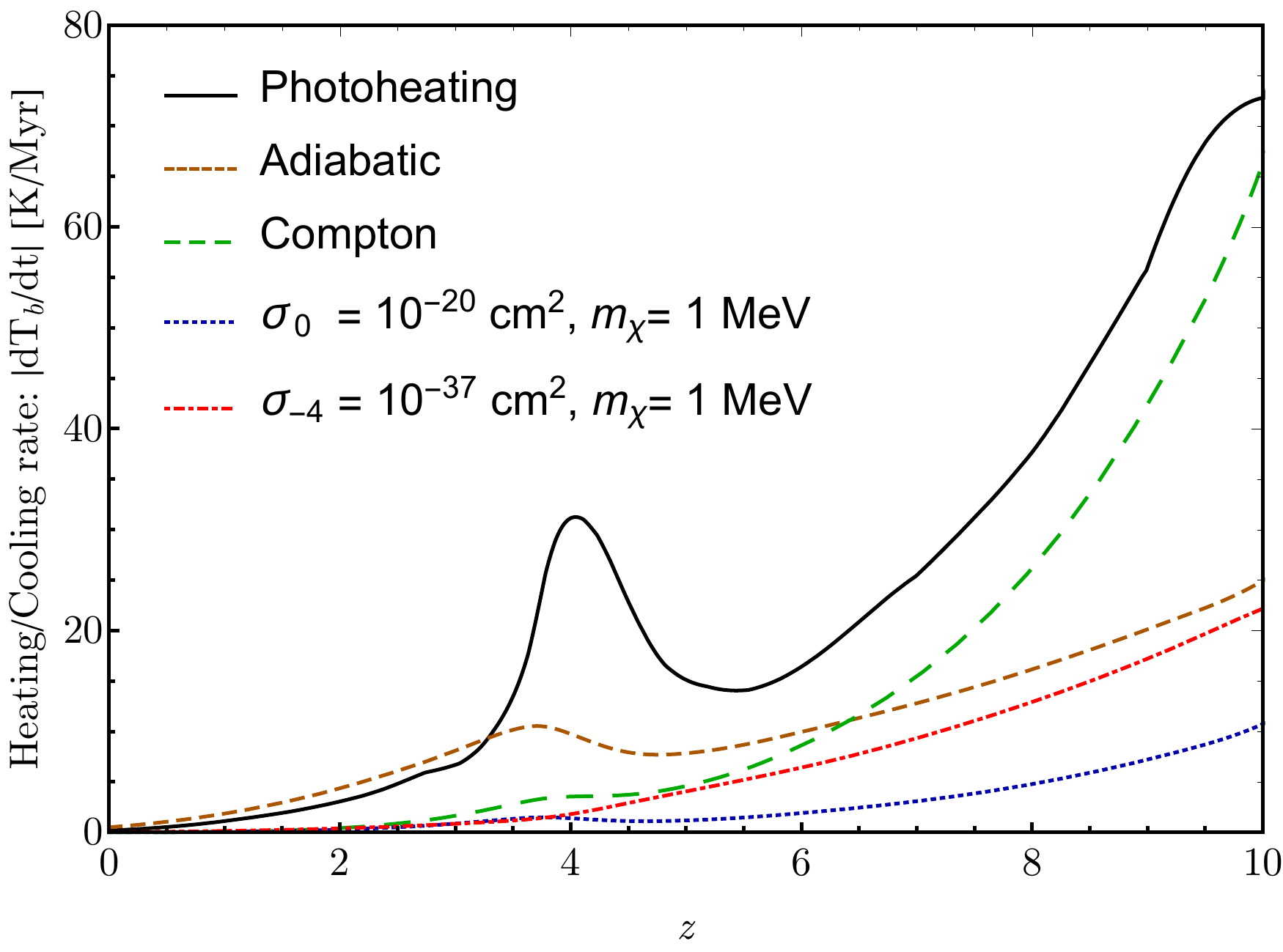}
	\caption{
		Heating and cooling rates from Eq.~\eqref{eq:TbdotIGM} in K/Myr as a function of $z$ for the average-density IGM. 
		In solid-black we show the photoheating term $\dot Q_{\rm ph}$, in dashed-brown the adiabatic term $- 2 H T_b$, in  long-dashed--green the Compton term $\Gamma_C (T_\gamma-T_b)$. We also show the cooling for two different cases of baryons interacting with a 1-MeV dark-matter particle, in dotted-blue for $n=0$ with $\sigma_0= 10^{-20}$ cm$^2$, and in dash-dotted--red for $n=-4$ and $\sigma_{-4}= 10^{-37}$ cm$^2$.
	}
	\label{fig:Rates}
\end{figure*}

However, at $z \sim 4$ HeII gets reionized, likely due to X-ray emission from quasars~\cite{Sokasian:2001xh,Madau:2015cga}, and loses its second electron. This causes an increase in photoheating and thus raises the gas temperature~\cite{Hui:1997dp,Abel:1999vj}.
Given the uncertainties in this process we will not attempt to model it, and instead focus on the redshift range $4<z<7$.
During this era, the (average-density) baryonic and dark-matter temperatures---including interactions---evolve according to \cite{Sanderbeck:2015bba,Lidz:2014jxa}
\ba
\dot T_b &= - 2 H T_b + \Gamma_C (T_\gamma-T_b) + \dfrac{2}{3}\Gamma_{b,\chi} (T_\chi-T_b) + \dfrac{2}{3} \dot Q_{\rm ph}, \label{eq:TbdotIGM} \nonumber
 \\
\dot T_\chi &= - 2 H T_\chi + \dfrac{2}{3}  \Gamma_{\chi,b} (T_b-T_\chi),
\end{align}
where overdots represent derivatives with respect to proper time, and we have defined $\Gamma_{\chi,b}\equiv\Gamma_{b,\chi} n_b/n_\chi$. In Eq.~\eqref{eq:TbdotIGM} the Compton thermalization rate is~\cite{Ma:1995ey}
\be
\Gamma_C = \dfrac{8 \mu_b \rho_\gamma n_e\sigma_T}{3 m_e \rho_b} ,
\label{eq:GammaC}
\ee
and $\dot Q_{ph}$ is the photoheating term arising from both H and He ionizations. We compute this term as
\be
\dot Q_{\rm ph} = \dfrac{1}{2} \, \dfrac{1}{1 + \chi_{\rm He}} \sum_X f_{X} (z) \Gamma_{h,X},
\ee 
where 
$\chi_{\rm He} \equiv n_{\rm He}/n_{\rm H} = 
Y_{\rm He}m_H/[(1 - Y_{\rm He})m_{\rm He}]$,
and $Y_{\rm He} = 0.24$ is the helium mass fraction, the index
$X$ runs over $\{\rm HI,HeI,HeII\}$ and $f_X$ is their fraction, and $\Gamma_{h,X}$ are the photoheating rates, obtained from the CUBA\footnote{\tt http://www.ucolick.org/$\sim$pmadau/CUBA/}
code~\cite{Haardt:2001zf,Haardt:2011xv} by using the average ionization fractions from Ref.~\cite{Puchwein:2014zsa}. 
We note that dark-matter--baryon interactions can also change the ionization history, although 
for the purposes of obtaining an order-of-magnitude constraint we ignore this effect.

Figure~\ref{fig:Rates} shows the redshift dependence of the different terms of Eq.~\eqref{eq:TbdotIGM} for the standard IGM temperature evolution. 
It is clear that at mean density the photoheating term dominates and is compensated by both adiabatic and Compton cooling to reach thermal equilibrium. 
Moreover, the effects of He reionization are imprinted in the photoheating  bump at $z\sim 4$. 
We also show the cooling due to baryon interactions with dark-matter particles of mass $m_\chi=$ 1 MeV for two cases: $n=0$ with a cross section of $\sigma_0 = 10^{-20}$ cm$^2$, and $n=-4$ with $\sigma_{-4} = 10^{-37}$ cm$^2$, from where we can see that the effects of interactions are more pronounced at earlier times, when the densities are higher.

\begin{figure*}[hbtp!]
		\centering
	\includegraphics[width=0.9\textwidth]{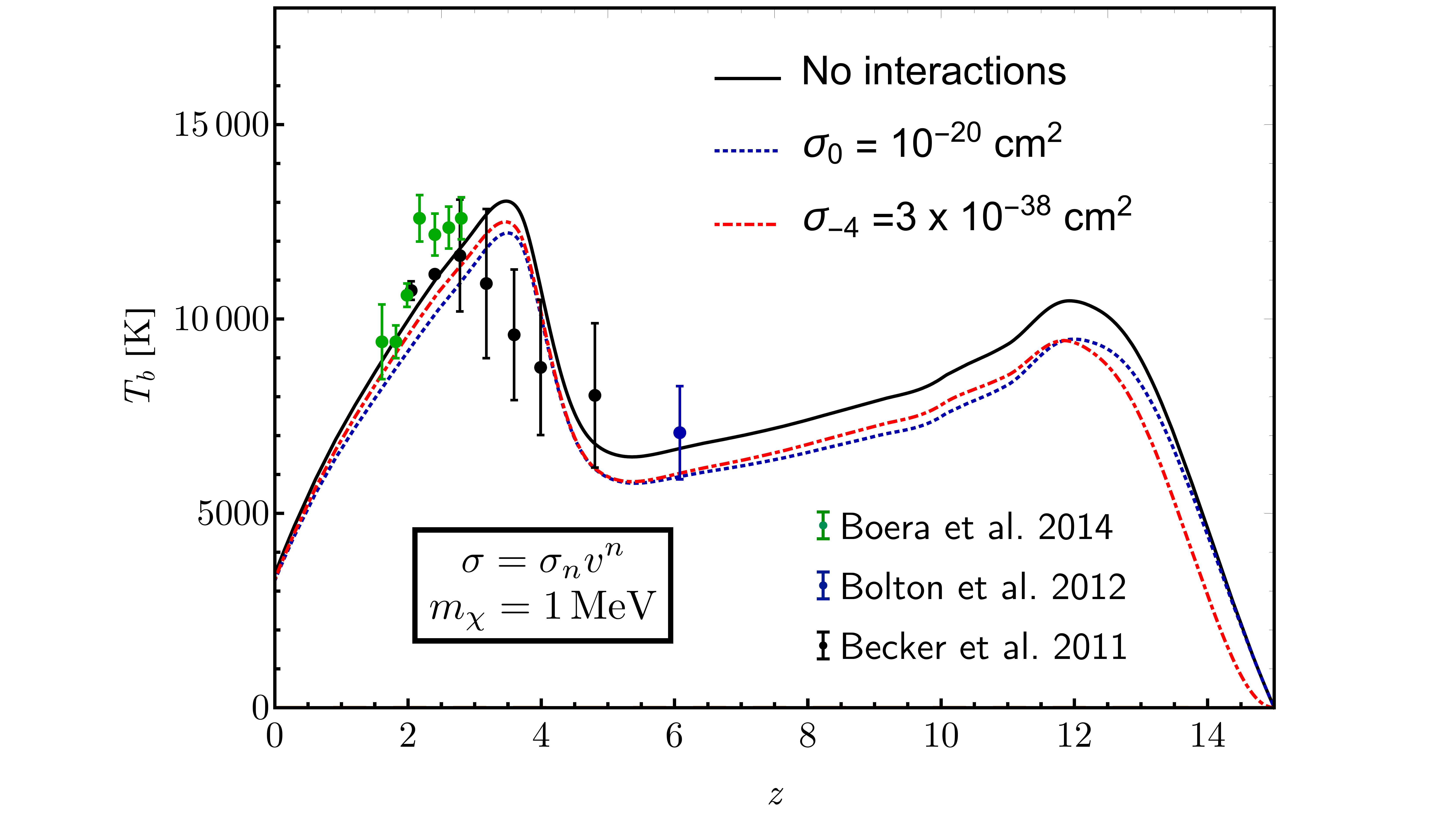}
	\caption{
		Average-density baryon temperature as a function of redshift obtained by integrating Eq.\,(\ref{eq:TbdotIGM}). The different lines represent baryon temperatures,
		in solid-black for the standard case without interactions,
		in dotted-blue and dot-dashed--red for a 1-MeV  interacting dark-matter particle, with $\sigma_0 = 10^{-20}$ cm$^2$ ($n=0$)  and with $\sigma_{-4}= 3\times 10^{-38}$ cm$^2$ ($n=-4$), respectively.
		We also show the datapoints from Refs~\cite{Boera:2014sia,Becker2011,Bolton2012} for comparison.
	}
	\label{fig:Tb}
\end{figure*}

In order to find the gas temperature $T_b$ as a function of redshift
we integrate Eq.\,(\ref{eq:TbdotIGM}), starting at $z_i=15$ with initially cold dark matter $T_\chi(z_i)=0$, 
and with a baryonic temperature given by $T_b (z_i) = T_\gamma (z_d)\,(1+z_i)^2/(1+z_d)^2 $, where we have taken $z_d = 200$ to be
the redshift of thermal decoupling between baryons and the CMB photons.
In Fig.~\ref{fig:Tb} we show the ``standard" case with no interactions, as well as two cases where a 1-MeV dark-matter particle interacts with baryons, first with $n=0$ and $\sigma_0 = 10^{-20}$ cm$^2$, and then for $n=-4$ and $\sigma_{-4}= 3\times 10^{-38}$ cm$^2$.
We also show temperature measurements at different redshifts~\cite{Becker2011,Bolton2012,Boera:2014sia}, obtained from Lyman-$\alpha$ spectra, in Fig.~\ref{fig:Tb}.
These measurements have been marginalized over the slope of the temperature-density relation, in order to find the temperature at mean density~\cite{Puchwein:2014zsa}.

\section{Results}
\label{sec:Results}

In order to find the maximum amount of DM-baryon interactions allowed by the data we will follow a conservative approach and not include additional effects, such as Compton or free-free cooling~\cite{Lidz:2014jxa}. Moreover, since helium reionization is a complicated process, likely with large fluctuations between different gas patches, it is challenging to interpret the temperature datapoints at low redshift.
We will, therefore, restrict ourselves to the datapoints corresponding to $T_b=7100 \pm 1200$ K and $T_b=8000 \pm 1900$ K~\cite{Bolton2012,Becker2011}, both at 95 \% C.L., corresponding to redshifts $z=6.1$ and $z=4.8$, respectively, which should be clean of HeII reionization effects. 

\begin{figure}[hbtp!]
		\centering
	\includegraphics[width=0.49\textwidth]{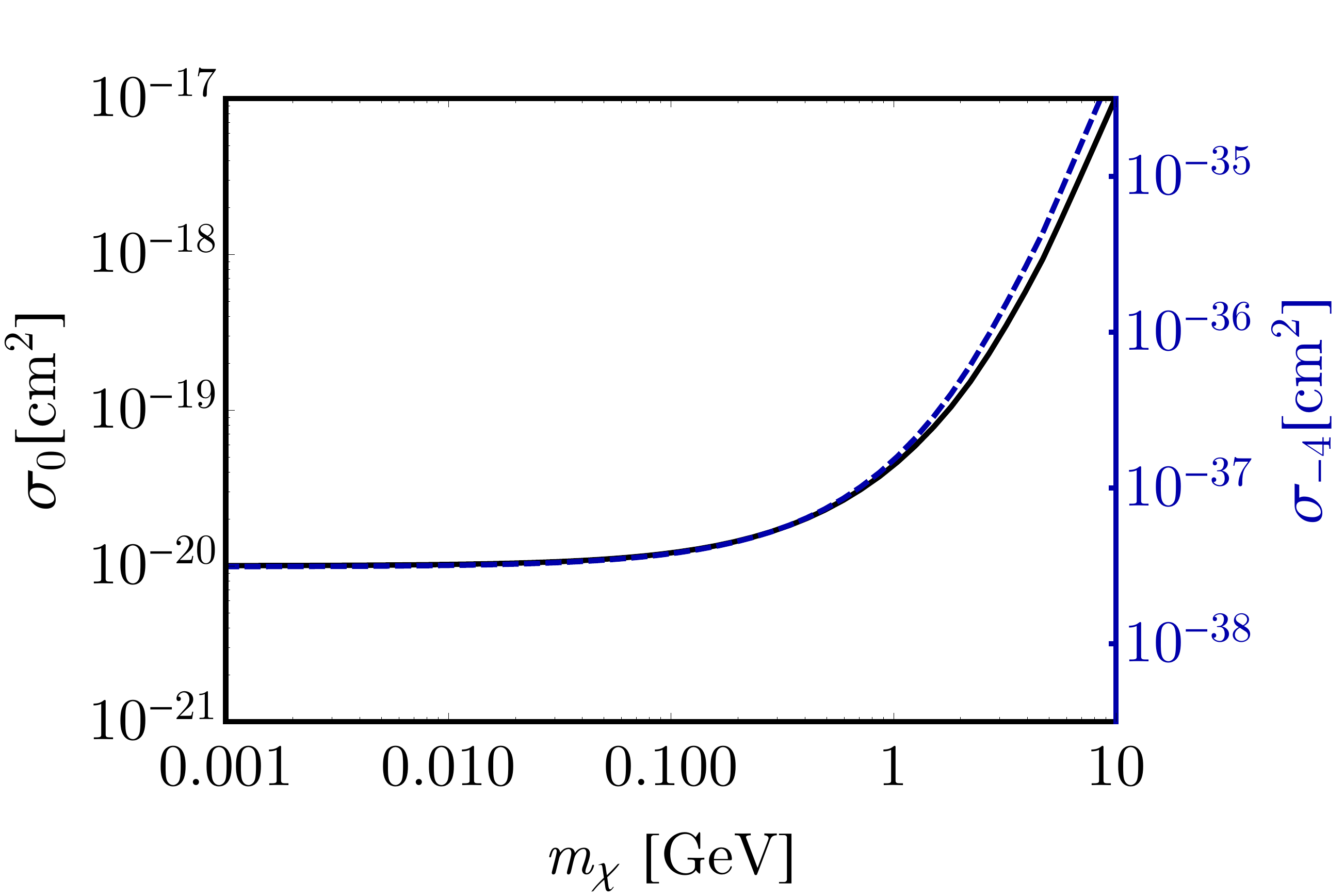}
	\caption{
		Maximum cross-section $\sigma_n$ allowed by the cooling of the baryons,
		We show the cases with $n=0$ and $n=-4$ as functions of the dark-matter particle mass $m_\chi$ in GeV.
	}
	\label{fig:sigmamin}
\end{figure}

We compute $T_b$ as a function of redshift for a set of dark-matter masses $m_\chi$ and interaction cross-sections $\sigma_n$ for $n=-4$ to $n=0$. Then,
by demanding that $T_b$ is within the error bars of both data points described above we arrive to constraints on the dark-matter--baryon cross section.
For simplicity, we will only consider the strongest constraint from both measurements and disregard the weaker one.
Figure~\ref{fig:sigmamin} shows the constraints for $n=0$ and $n=-4$, as functions of $m_\chi$.
In this Figure we can see that for low dark-matter mass the constraints are independent of $m_\chi$, as expected, since from Eq.~\eqref{eq:Qdotb} $\dot Q_b \propto m_\chi^0$ for $m_\chi\ll m_b \sim 1$ GeV. The IGM temperature measurements rapidly lose constraining power for $m_\chi\gtrsim 1$ GeV. This is due to the decrease in the dark-matter number density, which
via equipartition leads to the baryons having to share energy with fewer particles and thus a more modest decrease in the IGM temperature.
We show the maximum allowed value of the cross section---in the $m_\chi\ll m_p$ limit---in Table~\ref{tab:sigma}, for different $n$.

	\begin{table}[h]
			\centering
		\begin{tabular}{| c | c |}
			\hline
			n & $\sigma_n$ (cm$^2$) \\ \hline \hline
			$-4$ \ \ & $3.1 \times 10^{-38}$ \\
			$-2$ \ \ & $2.5 \times 10^{-29}$ \\
			$-1$ \ \ & $5.3 \times 10^{-25}$ \\
			$0$  &  $1.0 \times 10^{-20}$  \\     
			\hline
		\end{tabular}
		\caption{Maximum cross section allowed for interactions between baryons and  dark-matter particles with mass $m_\chi \ll 1$ GeV, parametrized as in Eq.~\eqref{eq:sigman}. The constraints arise from the gas-temperature measurements of Refs.~\cite{Bolton2012,Becker2011}, both at 95\%~C.L.}
		\label{tab:sigma}
	\end{table}

Since the baryon temperature is roughly constant during the redshift range of interest we can estimate their thermal velocity as $v_{\rm th} = \sqrt{3 T_b/m_b} \approx 10$ km/s. We can then translate our constraints $\sigma_n$ for each $n$ model in Table~\ref{tab:sigma} to a total dark-matter--baryon scattering cross section $\sigma = \sigma_n v_{\rm th}^{-n}$, which enables us to write 
\be
\sigma < 10^{-20} \,\rm cm^2
\ee 
for $m_\chi\leq 1$ GeV, independently of $n$.
This shows that the specific form of the interactions chosen does not alter the constraints dramatically.

Before concluding we will detail some additional studies we have performed.

\subsection{High-Redshift Heating}

As a check, we have solved the temperature evolution from Eq.~\eqref{eq:TbdotIGM} starting at $z_i = 2000$ at the limits $\sigma_n$ obtained above, to explore the behavior prior to our initial redshift $z_i=15$.
We find that Compton heating dominates prior to recombination, binding gas temperature to the photon temperature,
and thus no constraints from the gas-temperature can be inferred at this early epoch. 
However, it can be shown that dark-matter--baryon interactions with $n \geq 1$ produce baryonic cooling growing more rapidly than Compton heating at large $z$.
To avoid evolving Eq.~\eqref{eq:TbdotIGM} from some extremely large redshift we only report results for $n\leq 0$.

Moreover, we have checked that our results do not depend on the initial conditions $T_b$ and $T_\chi$ at $z_i = 15$ as long as both temperatures are small ($T_{b},T_\chi\lesssim 100$ K). So even if dark matter and baryons were in thermal equilibrium before $z \sim 15$, the fact that reionization heats baryons causes net heating of dark matter and cooling of the baryons.
This can be viewed as photoheating erasing all information encoded in the IGM temperature prior to the epoch of reionization.

\subsection{Dependence on the Photoheating Rate}

The effects of the dark-matter--induced cooling on baryons can be degenerate with the modeling of photoheating.
In Ref.~\cite{Sanderbeck:2015bba} it was shown that the reionization redshift $z_{\rm rei}$ does not have a large impact on photoheating, as long as $z_{\rm rei} \gtrsim 7$, as indicated by CMB data \cite{Adam:2016hgk}.
However, it was also shown that at $z\sim 5$ hardening the spectrum of the ionizing radiation can reduce the gas temperature. This could be confused with cooling arising from dark-matter--baryon interactions.
There are two possible avenues to breaking this degeneracy between photoheating and interactions.

First, future temperature measurements might be able to distinguish the different redshift evolution of photoheating and cooling induced by interactions, as shown in Fig.~\ref{fig:Rates}.
This would, however, require detailed understanding of photoheating---and thus emission spectra---across a wide redshift range.

Second, throughout this work we have computed the gas temperature at mean density, whereas it is well known that higher-density gas environments tend to have higher temperatures~\cite{Hui:1997dp}. 
The ``equation of state" of the IGM relates its temperature $T_{\rm IGM}$ to its density contrast $\Delta\equiv\rho/\bar\rho$ with respect to the average density $\bar\rho$. 
The temperature of the IGM at $z \leq 7$ is determined by the equilibrium between adiabatic cooling and photoheating. It is thus expected that denser regions, which would produce more photoheating whilst also supressing adiabatic expansion, would have larger temperatures. 
This is usually represented through the temperature-density relation
\be
T_{\rm IGM}(\Delta) = T_b \Delta^{\gamma-1}.
\ee
The interactions can change this relation, since denser regions also have larger dark-matter densities and thus cool the baryons further. We now study this effect.
For small overdensities ($\Delta\approx 1$),
one can find the evolution of the power-law index approximately as~\cite{Hui:1997dp}
\be
\dfrac{d(\gamma-1)}{da} = \dfrac{\Omega_M^{0.6}}{a} \left[ \dfrac 2 3 - (\gamma-1) \right] + \dfrac{2\dot Q_{\rm ph}}{3 a H T_b} \left [ 1-1.7 (\gamma-1) \right],
\ee
where we have assumed that the linear growth factor obeys $d\log D_+/d\log a = \Omega_M^{0.6}$, and the photoheating rate scales with temperature as $\dot Q_{\rm ph} \propto T_{\rm IGM}^{-0.7}$~\cite{Lidz:2014jxa}.
We can solve this equation with the initial condition $\gamma-1=0$ at $z_i=15$, and the fiducial reionization model presented in Section~\ref{sec:IGMTemperature}, to find that the power-law index asymptotes to a maximum value of
\be
\gamma_{\rm max}-1 = 0.596
\label{eq:gammamax0}
\ee
for $z \ll z_i$. This is in agreement with the result in Ref.~\cite{Hui:1997dp} for our fiducial value of $\Omega_M=0.3$.

Given the dark-matter-induced cooling rate from Eq.~\eqref{eq:Qdotb}, we can infer that
\be
\dot Q_{b,\chi} \propto T_{\rm IGM}^{(n+3)/2} \Delta_b,
\ee
for $T_\chi\approx 0$ and where we have assumed adiabatic initial conditions, i.e., $\Delta_\chi\propto \Delta_b$.
This will add a new term to the evolution of $\gamma-1$ as
\ba
\dfrac{d(\gamma-1)}{da} &= \dfrac{\Omega_M^{0.6}}{a} \left[ \dfrac 2 3 - (\gamma-1) \right]  + \dfrac{2\dot Q_{\rm ph}}{3a H T_b} \left [ 1-1.7 (\gamma-1) \right] \nonumber \\ 
&+ \dfrac{2\dot Q_{b,\chi}}{3a H T_b} \left [ 1- \dfrac{n+5}{2} (\gamma-1) \right].
\label{eq:gammadot}
\end{align}
We can thus find the change in $\gamma-1$ for different interaction indices $n$ and cross sections $\sigma_n$. 
We will only quote results for the asymptotic value at $z \ll z_i$, where we find
\ba
\gamma_{\rm max}-1 &\approx 0.596 + 0.03\ \, \times \left(\dfrac{\sigma_0}{10^{-20}\, \rm cm^2}\right) \quad {\rm for} \, n=0,\nonumber\\  
\gamma_{\rm max}-1 &\approx 0.596 + 0.02\ \, \times \left(\dfrac{\sigma_{-1}}{10^{-25}\, \rm cm^2}\right) \quad {\rm for} \, n=-1,\nonumber\\
\gamma_{\rm max}-1 &\approx 0.596 - 0.003\times \left(\dfrac{\sigma_{-2}}{10^{-29}\, \rm cm^2}\right) \quad {\rm for} \, n=-2,\nonumber\\
\gamma_{\rm max}-1 &\approx 0.596 - 0.03\ \, \times \left(\dfrac{\sigma_{-4}}{10^{-38}\, \rm cm^2}\right) \quad {\rm for} \, n=-4.
\label{eq:gammaresult}
\end{align}
Here, the change of sign in the extra term for $n\leq-2$ arises because the $\dot Q_{b,\chi}$ contribution to Eq.~\eqref{eq:gammadot} is multiplied by a factor $(0.5 + 0.3 n)$, where we have used the value of $(\gamma_{\rm max}-1)$ from Eq.~\eqref{eq:gammamax0}, and thus changes signs between $n=-1$ and $n=-2$.

%Although current measurements of $\gamma-1$ are not precise enough to detect this shift in the logarithmic slope~\cite{}, we remain hopeful that future measurements of $\gamma-1$ might 

%The study of this effect, however, would require numerical simulations and goes beyond the scope of this work.

\subsection{Dependence on the Reionization History}

Throughout this work we have assumed the fiducial reionization history from Ref.~\cite{Puchwein:2014zsa}.
This is only an approximation because (\emph i) it neglects the change in the ionization fractions from interactions between dark matter and baryons, which can produce a change in the gas temperature, and
(\emph{ii}) assumes that reionization starts at an early time ($z\gtrsim12$), whereas all we know is that it has to be completed by $z \sim 7$~\cite{Adam:2016hgk}.
We now show a simple way to estimate how the latter effect changes our constraints.

From Fig.~\ref{fig:Tb} it is easy to see that the gas temperature is roughly constant for $4\lesssim z \lesssim 12$. This enables us to readily find an order-of-magnitude estimate of the constraints, by integrating the change in temperature
\be
\Delta T_b = \dfrac{2}{3} \int dt \,  \Gamma_{b,\chi}(T_\chi - T_b),
\ee
where we can now set $T_\chi =0$ and $T_b = 8000$ K, and plug the result from Eq.~\eqref{eq:Gammabchi} to find
\be
\left . \dfrac{\Delta T_b}{T_b} \right|_{z_i} \approx 5\times 10^{15} \left(\dfrac{\sigma}{\rm cm^2}\right) \left[ (1+z_{f})^{5/2} - (1+z_{i})^{5/2} \right],
\ee
which, for $z_{f} = 12$ (where the gas temperature stops growing) and $z_{i}=6$ (where our first datapoint lies), and demanding that $\Delta T_b/T_b\leq 0.1$, as current data has roughly 10\% precision, yields $\sigma\leq 4 \times 10^{-19}$ cm$^2$, in the same order of magnitude as the result derived in Sect.~\ref{sec:Results}, albeit with a simpler approach.
This shows how our constraints scale with the redshift $z_f$ at which gas first heats up.

\subsection{Spin Temperature}

Molecular gas in galaxies can provide an additional measurement of the gas temperature, as it exhibits spin transitions between rotational states that can be excited by the CMB. The ratio of populations of these levels has been used to measure the CMB temperature at $z \lesssim 3$ \cite{2011A&A...526L...7N}.
The molecular gas is clumped, so collisions between gas particles can also affect the population ratio, from where the gas temperature might be inferred \cite{2013A&A...551A.109M}.
This measurement, however, is challenging to interpret for dark-matter interactions due to many uncertainties, such as 
the unknown fraction of gas in molecular and atomic forms \cite{2015JPhCS.661a2013S}
as well as the effects of turbulence \cite{2013A&A...551A.109M}.
Given these uncertainties, and the fact that this redshift range $z\sim 2$ is already explored by Lyman-$\alpha$ measurements in Ref.~\cite{Dvorkin:2013cea}, we defer the analysis of this method to future work.

\section{Conclusions}
\label{sec:Conclusions}

We have shown that IGM temperature measurements from the Lyman-$\alpha$ forest open a new window into studying interactions between baryons and dark matter. 
The gas temperature is heated by ionizing photons and cooled down by adiabatic expansion and Compton scattering off CMB photons. This yields a roughly constant temperature throughout the redshift range $4\lesssim z \lesssim 12$.
Baryons would be cooled further by interactions with dark matter, and by comparing their predicted temperature with the measurements from Refs.~\cite{Becker2011,Bolton2012} we limited the scattering cross-section $\sigma$.

A lower photoheating rate can, however, mimic the effects of interactions, so
our constraints ought to be interpreted at the order-of-magnitude level.
Nonetheless, we have shown that interactions would also modify the equation of state of the IGM, by altering the value of $\gamma-1$. Therefore, precise measurements of this power-law index, especially of its asymptotic value $\gamma_{\rm max}-1$, will separate both effects.

%studying the effects of interactions in every gas parcel of a simulation, while simultaneously evolving the gas density, its temperature, as well as the dark-matter temperature,
%will break the degeneracy between photoheating and interactions.

We have only considered the effects of direct scattering between dark matter and baryons on the gas temperature.
A more detailed treatment of these interactions, including the effects of the new force mediator, as well as inelastic scattering, can be found for example in Refs.~\cite{Green:2017ybv,Bertuzzo:2017lwt}.

We can summarize our findings simply as: (\emph i) for dark-matter masses $m_\chi$ larger than the proton mass $m_p\approx 0.938$ GeV the effect of cooling on baryons is small, and thus only weak constraints can be derived; (\emph{ii}) for $m_\chi\lesssim 1$ GeV the constraints are independent of both $m_\chi$ and the interaction index $n$, and can be simply recast as 
\be
\sigma \equiv \sigma_n v^{-n} \leq 10^{-20} \rm \, cm^2.
\ee
This constraint complements the one from Ref.~\cite{Dvorkin:2013cea} of $\sigma < 10^{-22}$ cm$^2$ $\times \, (m_\chi$/GeV), valid only for $m_\chi\gg 1$ GeV;
as well as the constraint $\sigma < 10^{-24}$ cm$^2$ $\times$ (keV/$m_\chi)^{1/2}$ from Ref.~\cite{Ali-Haimoud:2015pwa}, valid only for $m_\chi\lesssim 100$ keV.
Moreover, ours is the first direct constraint on dark-matter--baryon scattering at $z \sim 5$, 
opening a new window to employ gas-temperature measurements to study the nature of dark matter.

\acknowledgments

We wish to thank I.\ Cholis and E.~Kovetz for pointed questions and discussions, and especially S.\ Bird for insightful comments on an earlier version of the manuscript.
JBM would like to thank the support of Harvard's ITC, where this collaboration was initiated. 
JBM was partially supported by NSF Grant No.~0244990, NASA NNX15AB18G, and the Simons Foundation.

\bibliography{DMb_bib}{}
\bibliographystyle{bibpreferences.bst}
	
\end{document}